\documentclass[%
reprint,
superscriptaddress,
nobibnotes,
 amsmath,
 amssymb,
 aps, 
 prl,
floatfix,
]{revtex4-2}

\usepackage{graphicx}
\usepackage{dcolumn}
\usepackage{bm}
\setcitestyle{numbers}
\usepackage[hidelinks]{hyperref}

\begin{document}

\preprint{APS/123-QED}

\title{Search for non-virialized axions with 3.3 -- 4.2 \textmu eV mass at selected resolving powers}

\author{A. T. Hipp}
\affiliation{University of Florida, Gainesville, Florida 32611, USA}
\author{A. Quiskamp}
\affiliation{University of Western Australia, Perth, Western Australia 6009, Australia}
\author{T. J. Caligiure}
\affiliation{University of Florida, Gainesville, Florida 32611, USA}
\author{J. R. Gleason}
\affiliation{University of Florida, Gainesville, Florida 32611, USA}
\author{Y. Han}
\affiliation{University of Florida, Gainesville, Florida 32611, USA}
\author{S. Jois}
\affiliation{University of Florida, Gainesville, Florida 32611, USA}
\author{P. Sikivie}
\affiliation{University of Florida, Gainesville, Florida 32611, USA}
\author{M. E. Solano}
\affiliation{University of Florida, Gainesville, Florida 32611, USA}
\author{N. S. Sullivan}
\affiliation{University of Florida, Gainesville, Florida 32611, USA}
\author{D. B. Tanner}
\affiliation{University of Florida, Gainesville, Florida 32611, USA}
\author{M. Goryachev}
\affiliation{University of Western Australia, Perth, Western Australia 6009, Australia}
\author{E. Hartman}
\affiliation{University of Western Australia, Perth, Western Australia 6009, Australia}
\author{M. E. Tobar}
\affiliation{University of Western Australia, Perth, Western Australia 6009, Australia}
\author{B. T. McAllister}
\affiliation{Swinburne University of Technology, John St, Hawthorn VIC 3122, Australia}
\author{L. D. Duffy}
\affiliation{Los Alamos National Laboratory, Los Alamos, New Mexico 87545, USA}
\author{T. Braine}
\affiliation{University of Washington, Seattle, Washington 98195, USA}
\author{E. Burns}
\affiliation{University of Washington, Seattle, Washington 98195, USA}
\author{R. Cervantes}
\affiliation{University of Washington, Seattle, Washington 98195, USA}
\author{N. Crisosto}
\email{Work was done prior to joining Amazon}
\affiliation{University of Washington, Seattle, Washington 98195, USA}
\affiliation{Currently at AWS Center for Quantum Computing, Pasadena, CA 91125, USA}
\author{C. Goodman}
\affiliation{University of Washington, Seattle, Washington 98195, USA}
\author{M. Guzzetti}
\affiliation{University of Washington, Seattle, Washington 98195, USA}
\author{C. Hanretty}
\affiliation{University of Washington, Seattle, Washington 98195, USA}
\author{S. Lee}
\affiliation{University of Washington, Seattle, Washington 98195, USA}
\author{H. Korandla}
\affiliation{University of Washington, Seattle, Washington 98195, USA}
\author{G. Leum}
\affiliation{University of Washington, Seattle, Washington 98195, USA}
\author{P. Mohapatra}
\affiliation{University of Washington, Seattle, Washington 98195, USA}
\author{T. Nitta}
\affiliation{University of Washington, Seattle, Washington 98195, USA}
\author{L. J Rosenberg}
\affiliation{University of Washington, Seattle, Washington 98195, USA}
\author{G. Rybka}
\affiliation{University of Washington, Seattle, Washington 98195, USA}
\author{J. Sinnis}
\affiliation{University of Washington, Seattle, Washington 98195, USA}
\author{D. Zhang}
\affiliation{University of Washington, Seattle, Washington 98195, USA}
\author{C. Bartram}
\affiliation{Stanford Linear Accelerator Center, Menlo Park, California, 94025, USA}
\author{T. A. Dyson}
\affiliation{Stanford University, Stanford, CA, 94305, USA}
\author{C. L. Kuo}
\affiliation{Stanford Linear Accelerator Center, Menlo Park, California, 94025, USA}
\affiliation{Stanford University, Stanford, CA, 94305, USA}
\author{S. Ruppert}
\affiliation{Stanford University, Stanford, CA, 94305, USA}
\author{M. O. Withers}
\affiliation{Stanford University, Stanford, CA, 94305, USA}
\author{M. H. Awida}
\affiliation{Fermi National Accelerator Laboratory, Batavia, Illinois 60510, USA}
\author{D. Bowring}
\affiliation{Fermi National Accelerator Laboratory, Batavia, Illinois 60510, USA}
\author{A. S. Chou}
\affiliation{Fermi National Accelerator Laboratory, Batavia, Illinois 60510, USA}
\author{M. Hollister}
\affiliation{Fermi National Accelerator Laboratory, Batavia, Illinois 60510, USA}
\author{S. Knirck}
\affiliation{Fermi National Accelerator Laboratory, Batavia, Illinois 60510, USA}
\author{A. Sonnenschein}
\affiliation{Fermi National Accelerator Laboratory, Batavia, Illinois 60510, USA}
\author{W. Wester}
\affiliation{Fermi National Accelerator Laboratory, Batavia, Illinois 60510, USA}
\author{J. Brodsky}
\affiliation{Lawrence Livermore National Laboratory, Livermore, California 94550, USA}
\author{G. Carosi}
\affiliation{Lawrence Livermore National Laboratory, Livermore, California 94550, USA}
\author{N. Du}
\affiliation{Lawrence Livermore National Laboratory, Livermore, California 94550, USA}
\author{N. Robertson}
\affiliation{Lawrence Livermore National Laboratory, Livermore, California 94550, USA}
\author{N. Woollett}
\affiliation{Lawrence Livermore National Laboratory, Livermore, California 94550, USA}
\author{C. Boutan}
\affiliation{Pacific Northwest National Laboratory, Richland, Washington 99354, USA}
\author{A. M. Jones}
\affiliation{Pacific Northwest National Laboratory, Richland, Washington 99354, USA}
\author{B. H. LaRoque}
\affiliation{Pacific Northwest National Laboratory, Richland, Washington 99354, USA}
\author{E. Lentz}
\affiliation{Pacific Northwest National Laboratory, Richland, Washington 99354, USA}
\author{N. E. Man}
\affiliation{Pacific Northwest National Laboratory, Richland, Washington 99354, USA}
\author{N. S. Oblath}
\affiliation{Pacific Northwest National Laboratory, Richland, Washington 99354, USA}
\author{M. S. Taubman}
\affiliation{Pacific Northwest National Laboratory, Richland, Washington 99354, USA}
\author{J. Yang}
\affiliation{Pacific Northwest National Laboratory, Richland, Washington 99354, USA}
\author{R. Khatiwada}
\affiliation{Fermi National Accelerator Laboratory, Batavia, Illinois 60510, USA}
\affiliation{Illinois Institute of Technology, Chicago, Illinois 60616, USA}
\author{John Clarke}
\affiliation{University of California, Berkeley, California 94720, USA}
\author{I. Siddiqi}
\affiliation{University of California, Berkeley, California 94720, USA}
\author{A. Agrawal}
\affiliation{University of Chicago, Chicago, Illinois 60637, USA}
\author{A. V. Dixit}
\affiliation{University of Chicago, Chicago, Illinois 60637, USA}
\author{E. J. Daw}
\affiliation{The University of Sheffield, Sheffield, S10 2TN, United Kingdom}
\author{M. G. Perry}
\affiliation{The University of Sheffield, Sheffield, S10 2TN, United Kingdom}
\author{J. H. Buckley}
\affiliation{Washington University, St. Louis, Missouri 63130, USA}
\author{C. Gaikwad}
\affiliation{Washington University, St. Louis, Missouri 63130, USA}
\author{J. Hoffman}
\affiliation{Washington University, St. Louis, Missouri 63130, USA}
\author{K. W. Murch}
\affiliation{Washington University, St. Louis, Missouri 63130, USA}
\author{J. Russell}
\affiliation{Washington University, St. Louis, Missouri 63130, USA}

\collaboration{ADMX Collaboration}\noaffiliation

\date{\today}

\begin{abstract}
The Axion Dark Matter eXperiment is sensitive to narrow axion flows, given axions compose a fraction of the dark matter with a non-negligible local density. Detecting these low-velocity dispersion flows requires a high spectral resolution and careful attention to the expected signal modulation due to Earth's motion. We report an exclusion on the local axion dark matter density in narrow flows of $\rho_a \gtrsim 0.03\,\mathrm{GeV/cm^3}$ and $\rho_a \gtrsim 0.004\,\mathrm{GeV/cm^3}$ for Dine-Fischler-Srednicki-Zhitnitski and Kim-Shifman-Vainshtein-Zakharov axion-photon couplings, respectively, over the mass range $3.3-4.2\,\mu\text{eV}$. Measurements were made at selected resolving powers to allow for a range of possible velocity dispersions.
\end{abstract}

\maketitle

 Despite comprising $\sim27\%$ of the universe's mass-energy density \cite{planckcollaborationPlanck2018Results2020}, dark matter remains one of physics's most intriguing and persistent mysteries. There is extensive evidence that most galaxies, including our own, are immersed in large, spherical, halos of dark matter particles \cite{zwickyRepublicationRedshiftExtragalactic2009c, rubin_rotational_1980, planckcollaborationPlanck2018Results2020}. One of the leading candidates proposed to explain dark matter is the axion \cite{wilczekProblemStrongInvariance1978,weinbergNewLightBoson1978, preskill_cosmology_1983,abbott_cosmological_1983,Dine:1982ah}, a hypothetical pseudoscalar particle that arose as a consequence of Peccei and Quinn's (PQ) solution to the strong CP Problem in quantum chromodynamics (QCD) \cite{pecceiConstraintsImposedCP1977,pecceiCPConservationPresence1977}. Two benchmark models in which the axion solves the strong CP problem are the Kim-Shifman-Vainshtein-Zakharov (KSVZ) \cite{kimWeakInteractionSingletStrong1979, shifmanCanConfinementEnsure1980} and Dine-Fischler-Srednicki-Zhitnitsky (DFSZ) \cite{Zhitnitsky_1980tq, dineSimpleSolutionStrong1981} models. They are parameterized by the dimensionless, model-dependent coupling constant $g_\gamma$, taking a value of $-0.97$ and $0.36$ for the KSVZ and DFSZ models, respectively.

Axions with a mass between $\sim 1-100\, \mu \text{eV}$ are particularly appealing as dark matter candidates because they can be produced non-thermally in the early universe in sufficient quantities to account for the observed dark matter density \cite{preskill_cosmology_1983,abbott_cosmological_1983}. Some models suggest that a fraction of the local axion dark matter density could exist as cold flows, with velocity dispersions as low as $\mathcal{O}(10)$ m/s \cite{Sikivie:1995dp,chakrabartyImplicationsTriangularFeatures2021, ohareAxionMiniclusterStreams2023b}. This Letter details the results of a search for such cold flows, focusing on axions with masses between $3.3-4.2$ $\mu$eV and a range of possible velocity dispersions. 

The axion haloscope, first proposed by Sikivie \cite{sikivieExperimentalTestsInvisible1983}, is a method to convert local halo axions into detectable photons through their expected coupling to two photons, with a coupling strength given by

\begin{equation}
g_{a\gamma\gamma} = \frac{\alpha g_\gamma}{\pi f_a}.
\end{equation}
Here $f_a$ is the PQ symmetry-breaking scale, which is directly related to the unknown axion mass $m_a$ through:

\begin{equation}
    m_a \approx 6\,\mu\text{eV} \left(\frac{10^{12}\,\text{ GeV}}{f_a}\right).
\end{equation}
By placing a resonant cavity in a strong DC magnetic field (a source of virtual photons), axions can be converted into detectable photons with a frequency given by 

\begin{equation}
f = \frac{m_a c^2}{h} + \frac{m_a}{2 h} \left({\vec{v}\cdot\vec{v}}\right),
\label{mass_energy}
\end{equation}
where $\vec{v}$ is the axion's velocity relative to the experiment and $h$ is Planck's constant. Tuning the resonant frequency of a cavity mode, whose electric field spatially overlaps with the applied magnetic field, to the axion frequency will resonantly enhance the axion-photon conversion power by the quality factor $Q$ of the cavity. The Axion Dark Matter Experiment (ADMX) \cite{asztalosLargescaleMicrowaveCavity2001,asztalosExperimentalConstraintsAxion2002,asztalosImprovedRfCavity2004,asztalosSQUIDBasedMicrowaveCavity2010a,duSearchInvisibleAxion2018c,braineExtendedSearchInvisible2020a,bartramSearchInvisibleAxion2021b} is one such direct detection experiment, and has achieved DFSZ-level sensitivity across the $2.7-4.1\,\mu\text{eV}$ axion mass range for virialized halo axions \cite{duSearchInvisibleAxion2018c,braineExtendedSearchInvisible2020a,bartramSearchInvisibleAxion2021b}. 

The ADMX experiment consists of a large $136\,\ell$ microwave cavity immersed in a 7.5 T DC magnetic field, maintained at cryogenic temperatures. Because the axion mass is unknown, the cavity's resonant frequency must be tunable to search for the axion over a wide range. The resonant frequency of the axion-sensitive TM$_{010}$ mode is tuned using two movable copper rods inside the cavity. Using the detector parameters for ADMX Run 1C, the expected power developed inside the cavity due to axion-photon conversion is given as \cite{sikivieExperimentalTestsInvisible1983}

\begin{equation}
	\begin{aligned}
& P_{\text {axion }} = 7.7 \times 10^{-23} \mathrm{~W}\left(\frac{V}{136\, \ell}\right)\left(\frac{B}{7.5 \mathrm{~T}}\right)^2\left(\frac{C}{0.4}\right) \\
& \times\left(\frac{g_\gamma}{0.36}\right)^2\left(\frac{\rho_a}{0.45\, \mathrm{ GeV} / \mathrm{cm^3}}\right)\left(\frac{f}{1 \,\mathrm{GHz}}\right)\left(\frac{Q_L}{80,000}\right).
\label{power}
\end{aligned}
\end{equation}
Here $V$ is the cavity volume, $B$ is the magnetic field strength, $C$ is a mode-dependent form factor which represents the spatial overlap between the cavity electric field and the applied magnetic field, $\rho_a = 0.45 \,\mathrm{GeV/cm^{3}}$ is the local dark matter density (assumed to be all axions in our analysis) \cite{readLocalDarkMatter2014a} and $Q_L$ is the loaded quality factor of the cavity.

This weak signal power is extracted from the cavity through a critically coupled antenna and then amplified using an ultralow noise Josephson Parametric Amplifier (JPA). The blackbody radiation noise due to the physical cavity temperature $T_\text{cav}$ and the noise added by the receiver chain $T_\text{amp}$ contribute to the total system noise $T_\text{sys}$, given by

\begin{equation}
	T_\text{sys} = T_\text{cav} + T_\text{amp}.
\end{equation}
The added noise from the first stage amplifier is the most critical since any added noise from further amplification stages is suppressed by the gain of this initial stage. A detailed description of the Run 1C experiment can be found in Ref. \cite{bartramSearchInvisibleAxion2021b}.

Most halo models assume the population of axions to be virialized -- an equilibrium distribution controlled by its gravitational field. Consequently, the lineshape is expected to be Maxwellian with a spectral width given by the velocity dispersion. These axions are nonrelativistic with velocity dispersion of $\mathcal{O}(10^{-3}c)$, comparable to the orbital velocity of the Sun around our Galaxy. The velocity dispersion determines the spectral width of the axion lineshape and results in an effective axion quality factor $Q_a\sim10^6$ for a Maxwellian distribution. The spectral linewidth is then expected to be $\mathcal{O}(1 \text{ kHz})$, given axions with a mass $m_a \approx 4 \text{ }\mu\text{eV}$. To search for these virialized axions, ADMX acquires data in the so-called ``medium resolution'' (MR) channel for a total measurement time of 100 seconds. The final MR spectrum has a width of 50 kHz and a spectral resolution of 100 Hz from $10^4$ averages of individual $10\,\text{ms}$ subspectra.

Axion populations with narrower velocity dispersions, denoted as $\delta v$, will have correspondingly narrower spectral widths, $\delta f$, where their relationship is given by

\begin{equation}
	\frac{\delta f}{f} \approx \frac{v\delta v}{c^2}.
	\label{dispersion}
\end{equation}
N-body simulations suggest the Maxwellian expected from the isothermal halo model may be narrower than previously predicted \cite{Lentz_2017}. In addition, some models suggest the existence of axion flows with narrow velocity dispersions, such as the caustic ring model and axion miniclusters \cite{sikivieCausticRingsDark1998, duffyHighResolutionSearch2006,duffyCausticRingModel2008,ohareAxionMiniclusterStreams2023b}. However, most axion haloscope experiments are designed to search for virialized axions, rendering them less sensitive to the narrow Maxwellian and wholly insensitive to narrow flows. In contrast, ADMX implements a ``high-resolution'' (HR) data stream in parallel to the MR channel to analyze power spectra with frequency resolutions as fine as 20 mHz.

\begin{figure*}[t]
\centering
    \begin{minipage}[c]{0.49\textwidth}
        \centering
        \includegraphics[width=\linewidth]{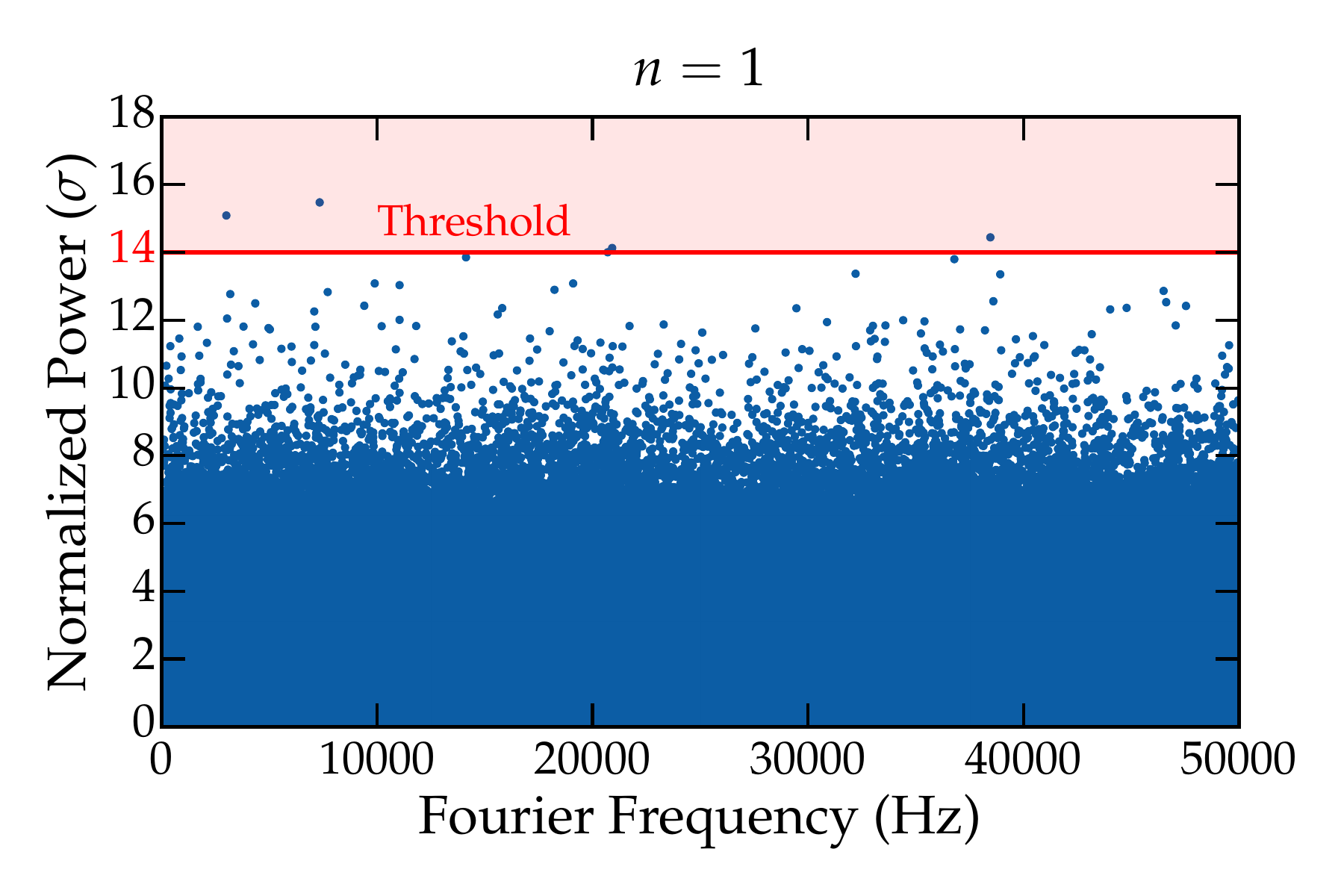}
    \end{minipage}
    \hfill 
    \begin{minipage}[c]{0.49\textwidth}
        \centering
        \includegraphics[width=\linewidth]{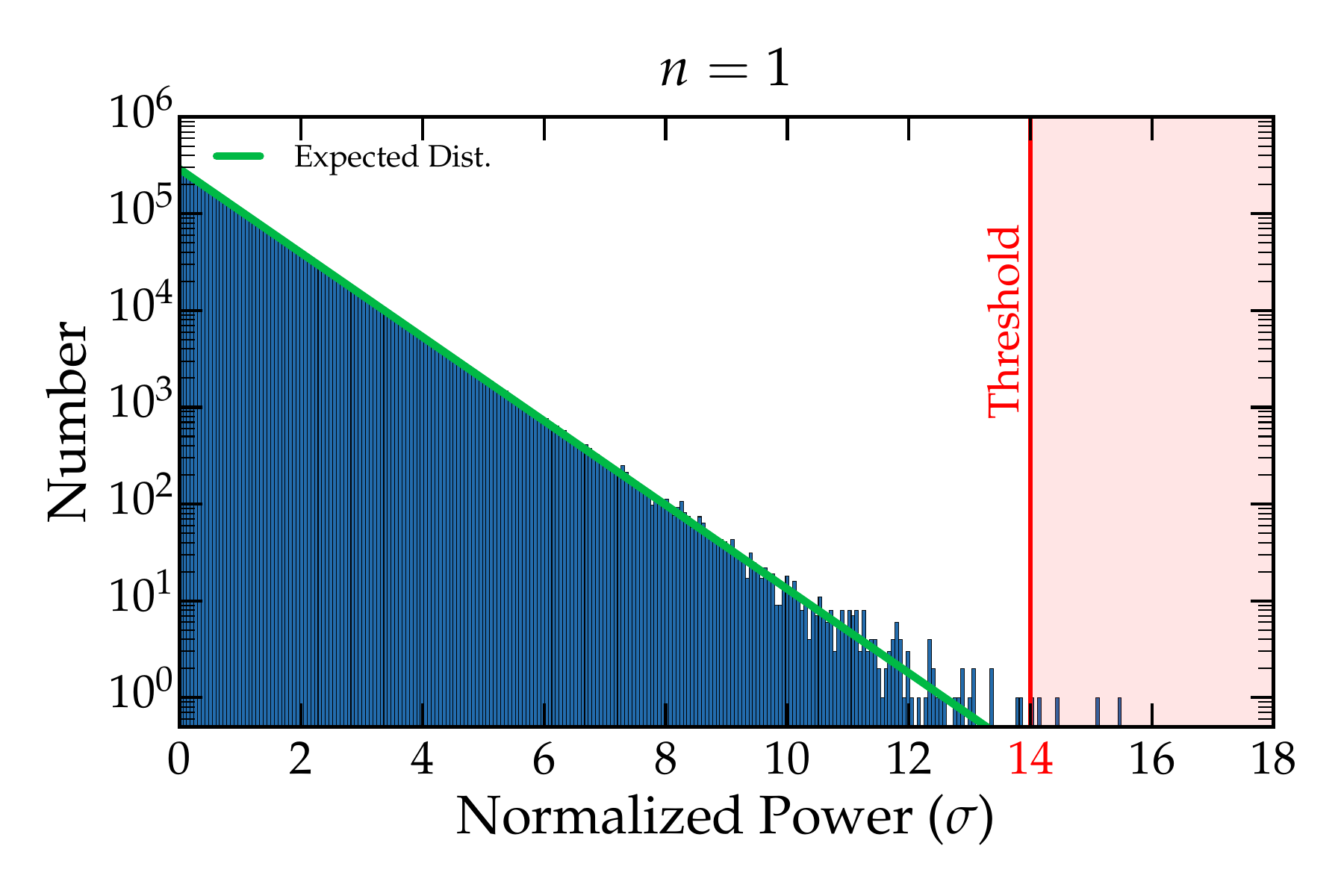}
    \end{minipage}
    \caption{\textbf{Left:} An example of a normalized power spectrum with $n=1$ is shown, illustrating how the normalized power varies across the $5,000,000$ Fourier frequency bins covering a range of $50\,\text{kHz}$. \textbf{Right:} A histogram of the normalized power for a typical spectrum with $n=1$. The expected Erlang distribution is shown in green, which for $n=1$ is an exponential distribution with a mean $\mu=1$ and standard deviation $\sigma=1$. The region shaded in red denotes the normalized power threshold of $P_T=14\sigma$.} 
    \label{1bin_dist}
\end{figure*}

Considering the potential variability in signal widths due to the uncertain velocity dispersions of specific cold flows in the caustic ring model or axion miniclusters, we conduct a high-resolution search across multiple frequency resolving powers. The maximum sensitivity to a given cold flow occurs when the bin width, $\Delta\nu$, matches the signal's full width at half the maximum power. A bin width of $10\,\text{mHz}$, for a 100-second digitization, is most sensitive to axions with velocity dispersion of $3\,\text{m/s}$, given a mean flow velocity magnitude of $v\approx 300\,\text{km/s}$ and a frequency $f=1\, \text{GHz}$. 

Coarser frequency resolutions are selected by subdividing the time series data into $n$ equal-length segments, where each is Fast-Fourier Transformed (FFT), its square magnitude calculated, and then normalized using a fifth-order polynomial fit to remove any spectral shape imparted by the receiver chain, then the set is summed. Resolution selection in the time domain is more robust against signals that become decoherent partway through the integration, compared to selecting coarser resolutions in the frequency domain by adding neighbouring bins \cite{hoskinsModulationSensitiveSearch2016}. By choosing $n=1,\, 10 \, \text{and} \,50$, we achieve sensitivity to flows across a wide range of velocity dispersions. These dispersions scale proportionally with $n$ relative to the native $10\,\text{mHz}$ bin width. 

\begin{figure*}[t]
\centering
    \begin{minipage}[c]{0.49\textwidth}
        \centering
        \includegraphics[width=\linewidth]{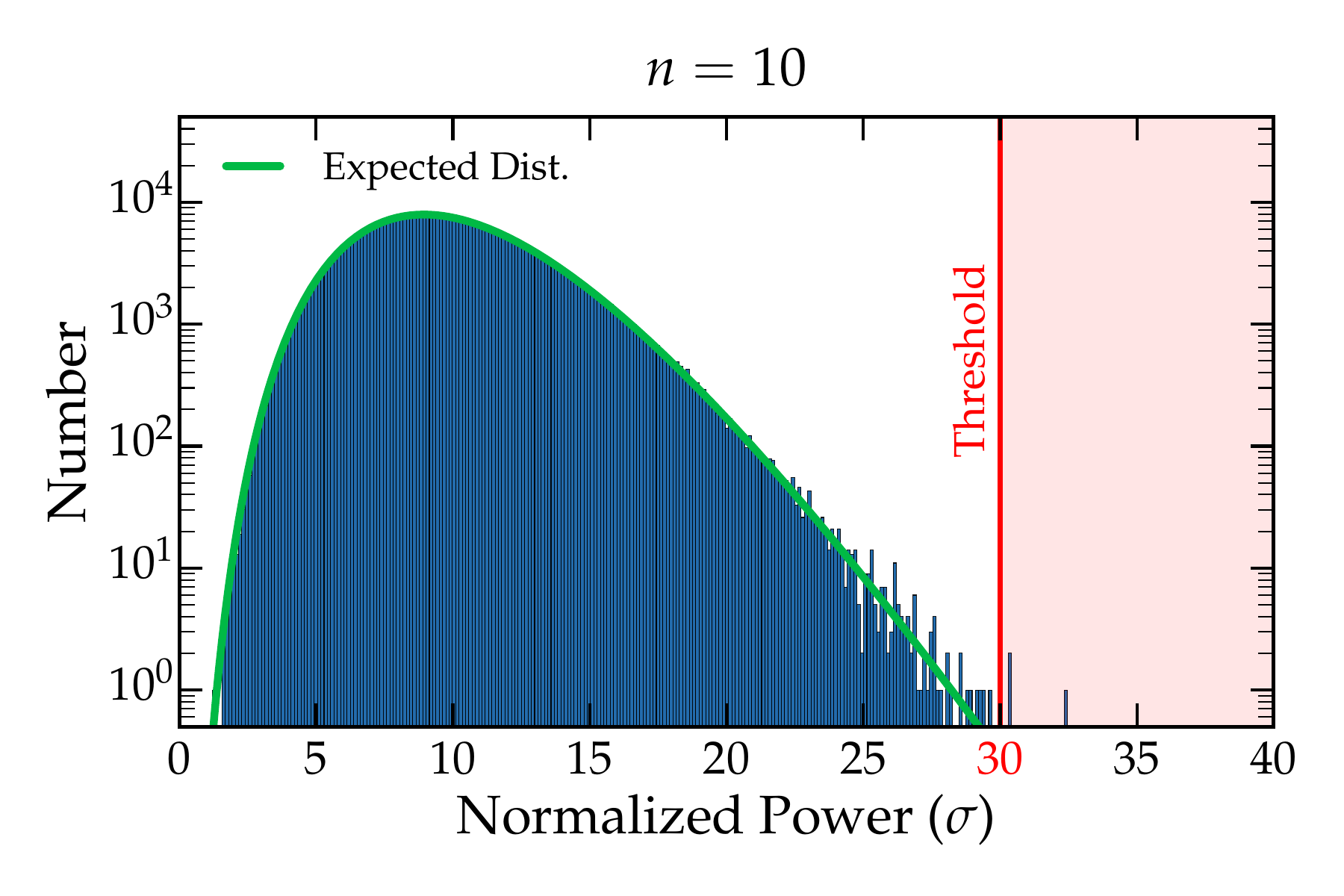}
    \end{minipage}
    \hfill 
    \begin{minipage}[c]{0.49\textwidth}
        \centering
        \includegraphics[width=\linewidth]{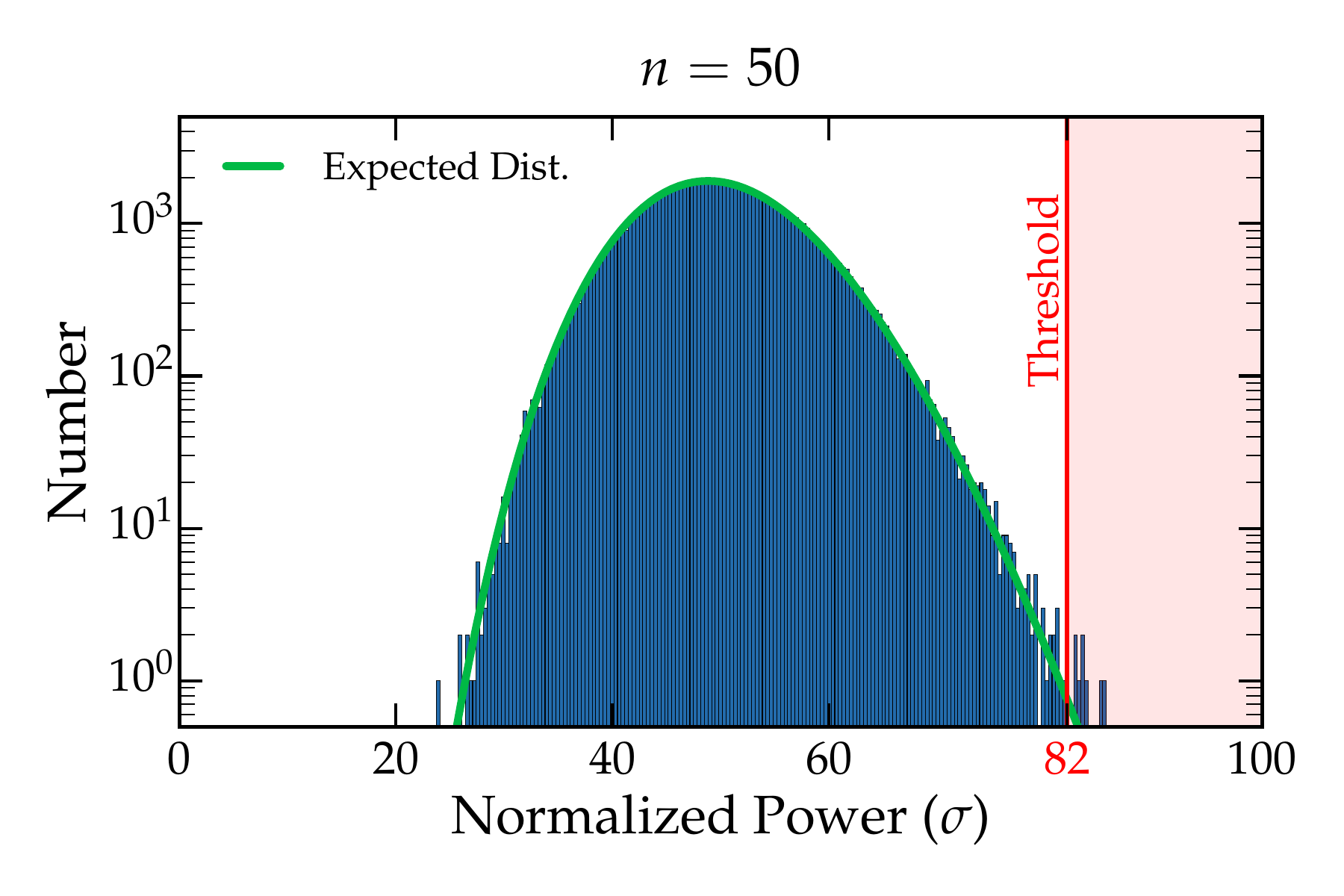}
    \end{minipage}
    \caption{A histogram of the normalized power for an example spectrum with $n=10$ and $n=50$ shown on the left and right, respectively. The expected Erlang distribution is shown in green with a mean (standard deviation) $\mu=10$ ($\sigma=\sqrt{10}$) and $\mu=50$ ($\sigma=\sqrt{50}$), respectively. The region shaded in red represents the chosen normalized power threshold of $P_T=30$ for the 10-bin search and $P_T=50$ for the 50-bin search.}
    \label{multi_bin_dist}
\end{figure*}

In general, the noise in the HR channel follows an Erlang distribution \cite{duffyHighResolutionSearch2006}, which is the sum of $n$ independent, identical exponential distributions. When $n=1$, the normalized power in a spectral bin follows an exponential distribution, with mean $\mu=n=1$ and standard deviation $\sigma=\sqrt{n}=1$ as shown in Fig. \ref{1bin_dist}. For the $n$-bin resolution, the probability $P$ of observing a noise power $W_n$ is given by \cite{duffyHighResolutionSearch2006}

\begin{equation}
	\frac{d P}{d W_n}=\frac{W_n^{n-1}}{(n-1)!} e^{-W_n}.
 \label{prob_dist}
\end{equation}
As spectra are subdivided and summed with increasing $n$, the probability distribution of the normalized power approaches a Gaussian distribution, as expected by the central limit theorem. This relationship can be seen in Fig. \ref{multi_bin_dist}. For reference, the MR channel corresponds to $n=10,000$ and thus is well approximated by a Gaussian distribution.

The threshold for a potential axion signal is based on the statistical properties of the normalized noise power. The power threshold $W_T$ in units of $\sigma$ should be set high enough to exclude all noise-related fluctuations confidently. A threshold that produces $\mathcal{O}(1)$ statistical candidates accomplishes this. We set the threshold at $14 \sigma$, as that produces approximately $4.15$ candidates per digitization. The example histogram and normalized power spectrum shown in Fig. \ref{1bin_dist} shows four candidates exceeding $W_T$. The thresholds for the $n=10$-bin and $n=50$-bin resolutions are determined similarly, giving $W_T=30\sigma$ and $W_T=82\sigma$, respectively. 

Data quality cuts removed digitizations based on poor $Q_L$, $T_{\text{sys}}$, and antenna coupling $\beta$, consistent with previous analyses \cite{ram_thesis, bartramNonvirializedAxionSearch2024, hoskinsModulationSensitiveSearch2016}. A calibration procedure to facilitate the MR analysis is the deliberate injection of synthetic axion signals through the weakly coupled port of the cavity. These synthetic signals are designed to mimic a Maxwellian distribution, serving as a control to test the sensitivity and analysis pipeline of the MR search. We removed all synthetic axion signals and any other contamination due to radio frequency interference (RFI) from the final data set. Additionally, candidates outside the bandwidth of the cavity, defined by the interval $f_0 \pm \frac{f_0}{1.8 Q_L}$, where $f_0$ is the resonant frequency, are removed.

The resulting, cleaned list of $\mathcal{O}(500,000)$ candidates is then analyzed for potential genuine axion signals. We expect an axion at a given frequency to remain persistent across multiple digitizations. However, according to Eq. \eqref{mass_energy}, the relative motion of the ADMX detector with respect to the axion flow causes a Doppler shift in the detected frequency \cite{chakrabartyImplicationsTriangularFeatures2021}. The Earth's orbit around the Sun and the rotation of the detector on the Earth's surface introduce annual and diurnal frequency modulations of an axion signal. A detailed discussion of the expected modulation for cold flows in the caustic ring model can be found in \citeauthor{bartramNonvirializedAxionSearch2024} \cite{bartramNonvirializedAxionSearch2024}. Given the time separation between candidates, the maximum modulation possible was computed using axions with a purely radial velocity equal to that of the galactic escape velocity at Earth. Candidates with no eligible neighbours were removed. 

If a candidate appears across multiple digitizations within the maximum allowed frequency modulation, it is classified as ``persistent''. We then calculate its persistence ratio, $\Upsilon$, 

\begin{equation}{\label{eqtn:persist}}
    \Upsilon = \frac{\text{\# of times a candidate is persistent}}{\text{\# of times the candidate was scanned}},
\end{equation}
where the number of times scanned corresponds to how many digitizations the candidate's frequency was within the cavity bandwidth.  

\noindent We conservatively set a persistence threshold of $\Upsilon=0.3$. Candidates appearing less than $30\%$ of the time were removed. After this requirement, 26 candidates remained and were deemed persistent. Given the expectation that an axion signal will produce excess power during the entire 100-second digitization, we can divide a single scan into two 50-second halves and, once again, check whether a given candidate is persistent. After applying this additional ``transient'' cut, no candidates remained. 

Given that all candidates in the searched frequency range have been excluded, we now set a limit based on the expected axion power, as determined by the experimental parameters at each digitization. The level of axion exclusion power is determined by the candidate threshold $W_T$ and the RMS noise power given by 

\begin{equation}
	W_\mathrm{N} =  k_B  T_\mathrm{sys} \Delta\nu_{b},
\end{equation}

\begin{figure*}[t]
	\includegraphics[width=0.9\linewidth]{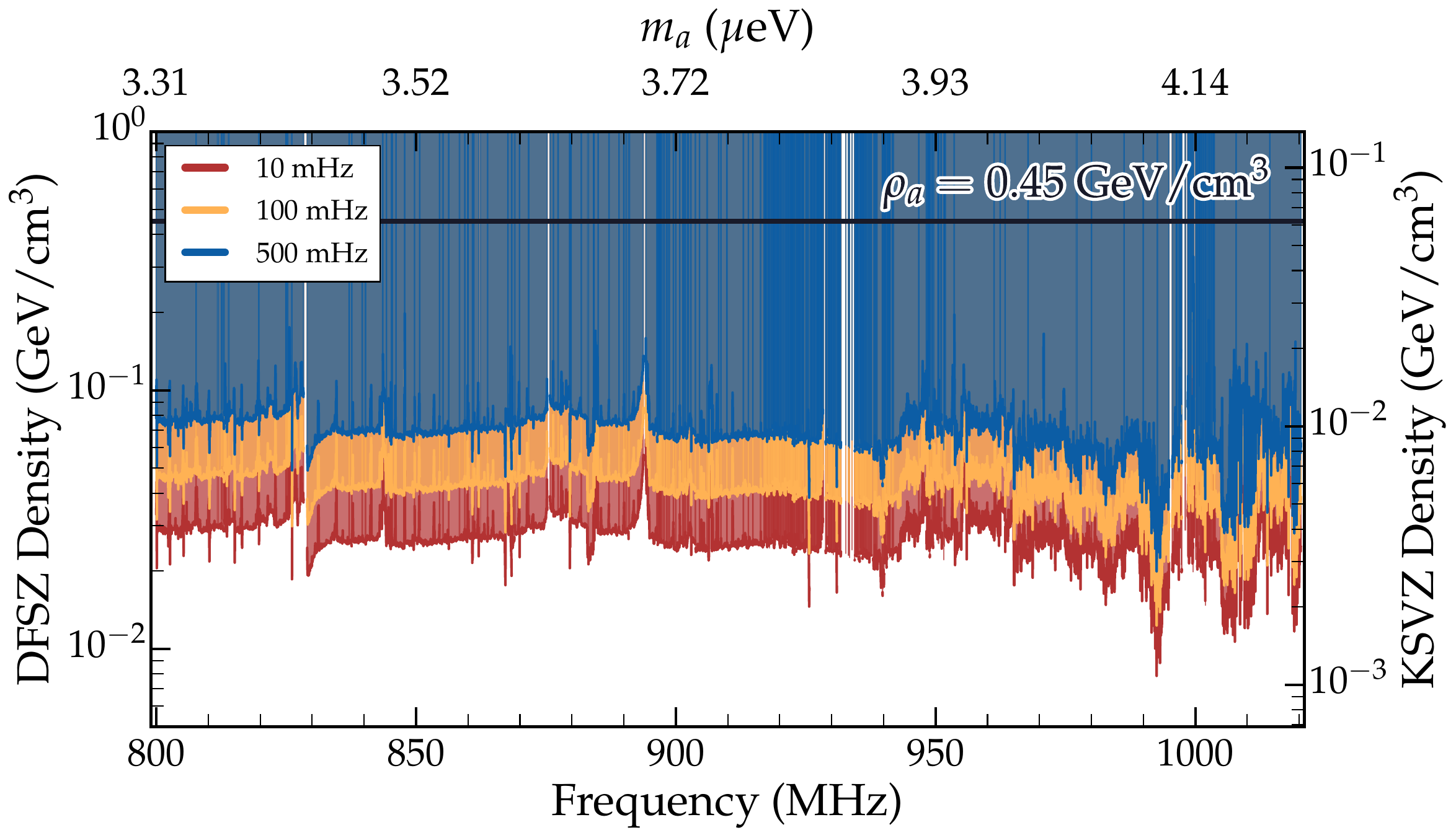}
 \caption{Exclusion limits on the local axion dark matter density assuming KSVZ (right axis) and DFSZ (left axis) axion-photon coupling for cold flows of varying velocity (frequency) dispersion at $95\%$ confidence interval. The horizontal line indicates the commonly assumed local dark matter density $\rho_a=0.45 \,\mathrm{GeV/cm^3}$. Narrow gaps in the plot represent mode crossings, RFI and regions that were not scanned with sufficient statistical significance.}
 \label{limits}
\end{figure*}

\noindent where $k_B$ is Boltzman's constant and $\Delta\nu_{b}$ is the spectral bin width. However, if an axion signal is present within a frequency bin, the observed power will be the sum of the axion-conversion power and the power from the fluctuating thermal background. Therefore, based on a specified confidence interval, we can attribute some amount of noise power to the background noise, which we denote as $W_B$. This results in a lower, effective threshold, given by $W_E=W_T-W_B$. For a $95\%$ confidence level, the power, in units of $\sigma$, that can be attributed to the background for a single scan is easily calculated using the probability distribution specified in Eq. \eqref{prob_dist}. 

The probability that the background passes the persistence cut 95\% of the time is given by the binomial distribution

\begin{equation}\label{success}
    \sum_{i=m}^{N} \binom{N}{i} F^{i}(1 - F)^{N-i} = 0.95,
\end{equation}
where $F$ is the probability the chosen order Erlang distribution has a power greater than $W_B$, $N$ is the total number of scans, and $m=\lceil \Upsilon N \rceil$ is the number of successes needed. Equation \eqref{success} was solved for $W_B$ at each frequency resolution in each frequency bin across the scan range. An axion signal with power greater than or equal to $W_E$ would have passed persistence 95\% of the time and thus detected, corresponding to a confidence level of 95\%. The value for $W_E$ is simply a function of the $N$-times a frequency bin was scanned, which varied over the search range. Limits were not computed for frequency ranges scanned zero or one times.

We now move to set limits on the local density of narrow flows, assuming KSVZ and DFSZ axion-photon coupling for each of the resolutions analyzed. This is different to other axion dark matter searches that instead set limits on the axion-photon coupling assuming a value for the dark matter density. As mentioned, this is done based on the expected axion power given by Eq. \eqref{power}. However, the detected power is reduced by several factors, including the antenna coupling $\beta$, the Lorentizan lineshape of the cavity mode and the power loss due to the discrete nature of FFTs. This loss occurs when the signal is not precisely centred within a bin, spreading signal power across neighbouring bins. The mean power loss due to the Lorentzian lineshape and bin-misalignment is calculated to be $\pi/4$ and $0.774$, respectively \cite{duffyHighResolutionSearch2006, hoskinsModulationSensitiveSearch2016}. Finally, limits on $\rho_a$ are calculated at a $95\%$ confidence interval assuming KSVZ and DFSZ axion-photon coupling. As outlined in \citeauthor{hoskinsModulationSensitiveSearch2016} \cite{hoskinsModulationSensitiveSearch2016}, we calculate a variance-weighted mean of contributing spectra. This weighting ensures those spectra with low $Q_L$ or high $T_{\text{sys}}$ contribute less to the density limits. 

Based on the ADXM Run 1C dataset, we exclude non-virialized DFSZ (KSVZ) axion flows with densities greater than $\sim 0.03 \,\mathrm{GeV/cm^3}$ ($\sim 0.004 \,\mathrm{GeV/cm^3}$) over the frequency range $800-1020\, \text{MHz}$ at the finest $20\,\text{mHz}$ bin resolving power. As shown in Fig. \ref{limits}, the limits at finer resolving power are more sensitive to $\rho_a$. However, this relies on the frequency dispersion of the expected axion signal matching the bin width. For example, the caustic ring model predicts a narrow flow with a frequency dispersion of $\mathcal{O}(250\,\text{mHz})$ and density of $12 \,\mathrm{GeV/cm^3}$, to pass through the Earth \cite{chakrabartyImplicationsTriangularFeatures2021}. The 50-bin search is maximally sensitive to this feature, thus this analysis rules out the caustic ring model in its current form. In general, the results presented can be used as a reference for constraining axion flow formation mechanisms.

This work was supported by the U.S. Department of Energy
through Grants No. DE-SC0009800, No. DESC0009723, No.
DE-SC0010296, No. DESC0010280, No. DE-SC0011665,
No. DEFG02-97ER41029, No. DEFG02-96ER40956, No.
DEAC52-07NA27344, No. DEC03-76SF00098, No. DE-SC0022148, and No. DESC0017987. Fermilab is a U.S. Department of Energy, Office of Science, HEP User Facility. Fermilab is managed by Fermi Research Alliance, LLC (FRA),
acting under Contract No. DE-AC02-07CH11359. Additional
support was provided by the Heising-Simons Foundation and
by the Lawrence Livermore National Laboratory and Pacific
Northwest National Laboratory LDRD offices. UWA participation is funded by the ARC Centre of Excellence for Engineered Quantum Systems, Grant No. CE170100009, Dark
Matter Particle Physics, Grant No. CE200100008, and Forrest
Research Foundation. The corresponding author is supported
by JSPS Overseas Research Fellowships No. 202060305.
LLNL Release No. LLNL-JRNL-853502.
\bibliographystyle{apsrev4-1}
\bibliography{hires_bib.bib}

\end{document}